\documentclass[12pt]{article}
\usepackage{amsmath}

\begin{document}
\providecommand{\e}[1]{\mathrm{e}^{#1}}

\begin{titlepage}

\begin{flushleft}
       \hfill                      {\tt hep-th/0103111}\\ 
       \hfill                       Imperial/TP/0-01/12 \\
       \hfill                       Napoli/10/2001 \\
       \hfill                       QMUL-PH-01-03 \\ 
\end{flushleft}
\vspace*{8mm}

\begin{center}
\textbf{\LARGE Holographic Renormalisation and Anomalies}\\
\vspace*{14mm}

{\large Jussi Kalkkinen$^a$, 
Dario Martelli$^b$ 
and Wolfgang M\"uck$^c$\footnote{E-mail addresses: 
\texttt{j.kalkkinen@ic.ac.uk} (J.~K.), 
\texttt{d.martelli@qmw.ac.uk} (D.~M.), \\
\texttt{wolfgang.mueck@na.infn.it} (W.~M.)}} \\
\vspace*{6mm}

\emph{${}^a$ Blackett Laboratory, Imperial College} \\
\emph{       Prince Consort Road, London SW7 2BZ, U.K.}
\vspace*{2mm}

\emph{${}^b$ Queen Mary \& Westfield College, University of London} \\
\emph{       Mile End Road, London E1 4NS, U.K.}
\vspace*{2mm}

\emph{${}^c$ Dipartimento di Scienze Fisiche, Universit\`a di
             Napoli ``Federico II''} \\
\emph{       Via Cintia, 80126 Napoli, Italy}

\vspace*{10mm}
\end{center}

\begin{abstract}
\noindent
The Weyl anomaly in the Holographic Renormalisation Group 
as implemented using Hamilton--Jacobi language is studied in detail.  
We investigate the breakdown of the descent equations in order to isolate   
the Weyl anomaly of the dual field theory close to the (UV) fixed point. 
We use the freedom of adding finite terms to the renormalised effective
action in order to bring the anomalies in the expected form.
We comment on different ways of describing the 
bare and renormalised schemes, and on possible interpretations of the descent
equations as describing the renormalisation group flow non-perturbatively.
We find that under suitable assumptions these relations may lead to a class of 
$c$-functions.
\end{abstract}

\end{titlepage}

\section{Introduction}
\label{intro}
The AdS/CFT correspondence \cite{Maldacena98,Witten98-1,Gubser98-1} 
has provided a remarkable example of holographic duality
\cite{'tHooft93,Susskind95}. According to this duality a theory 
containing gravity, for which it is at least very 
difficult to define local
observables, is described by a local quantum field theory
living on the boundary of space-time, and, \emph{vice versa}, physical
quantities of the boundary field theory can be obtained from the
bulk theory. Specifically, the AdS/CFT correspondence relates 
theories living on a $(d+1)$--dimensional anti-de Sitter (AdS) space to
conformal field theories (CFTs) living on the $d$--dimensional boundary.

Among the several proposed generalisations, the idea of implementing a 
holographic version of the renormalisation group has attracted considerable 
attention
\cite{Akhmedov98,Balasubrasmanian99,Freedman99a,DeWolfe00a,Alvarez99,Anselmi00a,Bianchi00,Hambli00,Graham00,Warner00,Evans00}.
This can be realised by considering string theory vacua in which 
some scalar field acquires a non-trivial dependence along one
direction, and the bulk space-time is no longer AdS. 
Such backgrounds are supposed to represent the running of the
couplings in terms of the so-called holographic coordinate,
which in turn is interpreted as an energy scale of the dual boundary
field theory. Several such supergravity solutions have been found
explicitly \cite{Girardello99a,Freedman00b,Brandhuber99} (see \cite{Gubser00-1}
for a review), and some correlation functions of the boundary
theories dual to these backgrounds have been computed 
\cite{Chepelev99-1,Rashkov99-3,Anselmi00a,DeWolfe00a,Arutyunov00a,Bianchi00}.  
Even if most of the solutions available display some kind of singularity 
in the IR, here we are interested in considering non-singular solutions 
that smoothly interpolate between two fixed point CFT's. We will 
attempt to find non-perturbative relations that should be valid along
these flows, and appropriately reduce to the fixed point results.

According to the general prescription of the holographic duality, one
must solve the equations of motion of the bulk theory and evaluate the
(suitably renormalised) on-shell action in order to obtain the 
generating functional of the boundary field theory. The only
exceptions to this effort are anomalies (or holographic one-point
functions), which can be evaluated without knowledge of the full bulk
solutions. Here, an analysis of the asymptotic behaviour of
the bulk fields towards the boundary is sufficient
\cite{Henningson98-2,deHaro00a,Schwimmer:2000cu,Manvelyan01}.

Another interesting approach to the study of the holographic
renormalisation group, using the Hamilton--Jacobi formulation of the
bulk theory, has been proposed by de Boer, Verlinde and
Verlinde (dBVV) \cite{deBoer00a,Verlinde00a,Verlinde00b,deBoer01a} and
studied further in
\cite{Campos00,Fukuma00a,Fukuma00b,Kalkkinen00,Nojiri00d,Kalkkinen01a}.  
The Hamilton equations of motion resemble the flow of couplings with the
cut-off scale, and a holographic beta function can be defined. 
Moreover, the boundary counter terms necessary to
renormalise the on-shell action are elegantly obtained using a
derivative expansion and solving the resulting descent
equations. Holographic anomalies occur whenever a descent equation 
breaks.

Despite the formal agreement of the results of dBVV's method with
those of the asymptotic analysis, the exact relation between dBVV's
interpretation of the holographic renormalisation group and the
standard implementation of holography has remained somewhat obscure.
In this paper, we shall try to elaborate this point by a systematic
study of dBVV's descent equations up to level four. Our results for
the anomalies in the vicinity of a given fixed point agree with those
of \cite{deHaro00a}, but are slightly more general. 
An interpretation of the descent equations along the full
renormalisation group flow proves more difficult. We find that a
holographic $c$-function for $d=2$ can be defined from the level two
descent equations, but such an interpretation becomes problematic in
higher dimensions. 

As a simple generic model, we consider the action for scalar fields
coupled to gravity in $d+1$ dimensions,\footnote{Our conventions for
the curvature tensor are 
$R^\mu{}_{\nu\rho\lambda}=\partial_\rho \Gamma^\mu{}_{\nu\lambda} +
\Gamma^\mu{}_{\rho\sigma}\Gamma^\sigma{}_{\nu\lambda}
-(\rho\leftrightarrow\lambda)$, $R_{\mu\nu}=R^\rho{}_{\mu\rho\nu}$.
Here we have adorned with a tilde quantities belonging to $(d+1)$ space
to distinguish them from those of the $d$-dimensional
hypersurfaces.} 
\begin{equation}
\label{intro:action}
  S = \int d^{d+1}x \sqrt{\tilde{g}} \left[ - \tilde{R} +\frac12
   \tilde{g}^{ab} \partial_a \phi^I G_{IJ}(\phi) \partial_b \phi^J -
  V(\phi) \right] + 2 \int d^d x \sqrt{g}\, H~.
\end{equation}
The second term on the right hand side of
eqn.\ \eqref{intro:action} is the Gibbons--Hawking boundary term
including the extrinsic curvature $H$ of the boundary.

Assuming that the potential $V$ has a local extremum at $\bar{\phi}$
with $V(\bar{\phi})>0$, there exists a solution to the classical equations of
motion of \eqref{intro:action} with $\phi=\bar{\phi}$ and
AdS geometry with cosmological constant $-2\Lambda= 
V(\bar{\phi}) = d(d-1)/l^2$. Such backgrounds give rise to boundary
CFTs, and the AdS/CFT correspondence has been
well-studied in the past years. In particular, the holographic Weyl
anomaly, derived first by Henningson and Skenderis
\cite{Henningson98-2}, is, for $d=2$ and $d=4$, respectively, 
\begin{align}
\label{intro:anomd2}
  \langle T \rangle &= -c_2 R~,\\
\label{intro:anomd4}
  \langle T \rangle &= c_4 I_4 - a_4 E_4~. 
\end{align}
($I_4$ and $E_4$ are the square of the Weyl tensor and the Euler
density in four dimensions, respectively.) The coefficients $c_2=l$,
$c_4=l^3/8$ and $a_4=l^3/8$ 
are identified as central charges of the boundary CFTs. Taking into account
the coupling with nontrivial scalar fields gives rise to additional terms,
that generically mix with curvature terms. Some of these terms 
were computed in \cite{Liu98-1} in the context of AdS/CFT. We will illustrate 
how such terms can be obtained for bare fields in the vicinity of the UV
fixed point by performing Hamilton--Jacobi analysis, and eventually
translated to renormalised quantities. We will also see that a
non-perturbative 
extrapolation of the results towards the IR fixed point proves more 
difficult.
 
One can use the following arguments \cite{Freedman00a} to identify UV
and IR fixed points. Scalar fluctuations around the constant scalar
background have a mass $m^2 = - 
\partial^2 V /\partial\bar{\phi}^2$ and act as sources of primary conformal 
operators of dimension $\Delta=d/2 + \sqrt{d^2/4+m^2l^2}$. Thus, at a
local minimum of $V$, tachyons are present in the bulk theory,
which act as sources of relevant operators ($\Delta<d$) in the
boundary CFT so that the field theory is unstable at this point (UV
fixed point). \emph{Vice versa}, at a local maximum of $V$, there are no 
tachyons in the bulk and only irrelevant operators in the CFT (IR
fixed point). For two neighbouring
extrema of $V$, we have $V_{\mathrm{UV}}<V_{\mathrm{IR}}$,
$l_{\mathrm{UV}}>l_{\mathrm{IR}}$, and thus generically
$c_{\mathrm{UV}}>c_{\mathrm{IR}}$ for the central charges. 

Let us conclude this section with an outline of the rest of the
paper. In Sec.\ \ref{review} we briefly review dBVV's method. The
descent equations are analysed in the vicinity of a fixed point in
Sec.\ \ref{perturb}, and the possible anomalies are derived. In Sec.\
\ref{global} we give conclusions and try to extend the meaning of the
descent equations to the full renormalisation group flow. The lengthy
expressions of level four are listed in the appendix.

\section{Hamilton--Jacobi Approach}
\label{review}
In this section, we briefly recall the Hamilton--Jacobi  
approach to the holographic renormalisation group. We shall put
special emphasis on the differences with the standard implementation
of holography in order to establish a suitable connection in later
sections. The reader is referred to
\cite{deBoer00a,Verlinde00a,Verlinde00b,Fukuma00a,Fukuma00b} for 
details of the method. 

The bulk field theory with the action \eqref{intro:action} is treated
in ADM formalism. The constraints in phase space are 
\begin{align}
\label{review:Hamilton}
  \mathcal{H} &= \pi^{\mu\nu} \pi_{\mu\nu} - \frac1{d-1} \pi^2 +\frac12
  \pi_I G^{IJ} \pi_J +R - \frac12 g^{\mu\nu} \partial_\mu \phi^I
  G_{IJ} \partial_\nu \phi^J + V \approx 0~, \\
\label{review:Hnu}
  \mathcal{H}^\nu &= \nabla_\mu \pi^{\mu\nu} -\frac12 \pi_I \nabla^\nu
  \phi^I \approx 0~.
\end{align}
We will choose lapse and shift functions such that the bulk metric is 
\begin{equation}
\label{review:metric}
  ds^2 = \tilde g_{ab} dx^a dx^b = dr^2 +g_{\mu\nu} dx^\mu dx^\nu
\end{equation}
($a,b= 0,\ldots,d$, $\mu,\nu=1,\ldots,d$) and the space-like foliation 
parameter is $x^0=r$, to which we will refer as the cut-off.
 
For a classical trajectory the on-shell action, henceforth denoted by
$S$, is a functional of the values $g_{\mu\nu}(r,x)$ and $\phi^I(r,x)$
at the cut-off (boundary values).
The canonical momenta at the cut-off are given by the 
functional derivative of $S$ with respect to the boundary data,
\begin{align}
\label{review:momenta}
  \pi_I &= \frac1{\sqrt{g}} \frac{\delta S}{\delta \phi^I}~, & 
  \pi^{\mu\nu} &= \frac1{\sqrt{g}} \frac{\delta S}{\delta g_{\mu\nu}}~,
\end{align}
so that both the constraints \eqref{review:Hamilton} and \eqref{review:Hnu}
become the Hamilton--Jacobi equations for $S$.
Moreover, (one half of) the Hamilton equations of motion yield the flow
equations, 
\begin{align}
\label{review:HJ1}
  \dot{\phi}^I &= G^{IJ} \pi_J~,\\
\label{review:HJ2}
  \dot{g}_{\mu\nu} &= 2\pi_{\mu\nu} -\frac2{d-1} g_{\mu\nu} \pi~.
\end{align}

According to the ``AdS/CFT master formula''
\cite{Gubser98-1,Witten98-1}, the on-shell action $S$ is identified
(up to possible local counter terms) with the generating functional of
the boundary field theory, where the (suitably rescaled) boundary data
play the role of the sources coupling to the boundary field theory operators. 
De~Boer, Verlinde and Verlinde \cite{deBoer00a} proposed to
use a derivative expansion to systematically determine the counter
terms and wrote the on-shell action $S$ as 
\begin{equation}
\label{review:S}
  S = [S]_0 + [S]_2 + [S]_4 +\cdots + \Gamma~,
\end{equation}
where 
\begin{align}
\label{review:S0}
  [S]_0 &= \int d^dx\sqrt{g}\, U(\phi)~,\\ 
\label{review:S2}
  [S]_2 &= \int d^dx\sqrt{g} \left[\Phi(\phi) R + \frac12
  g^{\mu\nu} \partial_\mu \phi^I M_{IJ}(\phi) \partial_\nu \phi^J
  \right], \quad \text{etc.,}
\end{align} 
are the local counter terms involving $0,2,\ldots$ derivatives.
$\Gamma$, which is generically non-local, is identified with the
generating functional of the boundary field theory, so that
\begin{align}
\label{review:T}
\langle\mathcal{O}_I \rangle &= \frac1{\sqrt{g}} \frac{\delta
\Gamma}{\delta \phi^I}~, &
\langle T^{\mu\nu} \rangle &= \frac2{\sqrt{g}} \frac{\delta
\Gamma}{\delta g_{\mu\nu}}~.
\end{align}
Such relations hold both for the bare $\phi(x,r)$  and renormalised 
$\hat\phi(x)$ fields \cite{deBoer00a}. We define the bare fields as 
solutions of full bulk equations of motion with the asymptotic 
boundary conditions at $r\to \infty$,
\begin{align}
\label{review:asym1}
  g_{\mu\nu}(x,r) &= \e{2r/l} \hat{g}_{\mu\nu}(x) + \cdots~,\\
\label{review:asym2}
  \phi(x,r) &= \e{(\Delta-d)r/l} \hat{\phi}(x) +\cdots~.
\end{align}
The fields $\hat{g}_{\mu\nu}$ and $\hat{\phi}$ are the boundary
field theory sources coupling to the energy momentum operator,
$\hat{T}^{\mu\nu}$, and some scalar operator, $\hat{\mathcal{O}}$,
respectively. 
Because in the Hamilton--Jacobi theory one naturally works in terms of 
bare couplings, we need to give a procedure for finding renormalised  
correlators. As the two sets of 
fields are related by the equations of motion, one finds
\begin{equation}
\label{jacob}
  \frac{\delta}{\delta \hat\phi(x)} = \frac{\delta \phi(x,r)}{\delta 
  \hat\phi(x)}\,  \frac{\delta}{\delta \phi(x,r)} \equiv 
  \mathcal{A}(x,r)\,  \frac{\delta}{\delta \phi(x,r)} 
\end{equation}
where $\mathcal{A}(x,r)$ acts as a bulk--boundary propagator. 
This operator is in general a complicated non-local operator, 
whose explicit form is not known. Nevertheless, its expansion for 
a single free scalar in fixed gravitational background is known 
\cite{deHaro00a}:
There, it turns out that apart from the scaling factor already explicit in 
(\ref{review:asym2}), $\mathcal{A}(x,r)$ is a 
series in the boundary d'Alembertians.
For the purposes of the present analysis, these corrections coming 
from the Jacobian (\ref{jacob}) will give rise to 
higher order derivative corrections, and it turns out that they can be 
consistently neglected (cf.~Sec.~\ref{perturb}).

Consistency of perturbation theory in the bulk field theory requires
$\Delta<d$, \emph{i.e.}, we are dealing with operators at the UV fixed
point.

The constraint \eqref{review:Hnu} is satisfied by each of the local
terms $[S]_{2n}$, whereas for $\Gamma$ it represents the Ward identity
for diffeomorphisms in the boundary field theory,\footnote{This turns out to 
be the case for scalar fields interacting with gravity, whereas the analogous
equation in the presence of interacting fermion fields obtains unusual
contributions \cite{Kalkkinen00}.}
\begin{equation}
\label{review:Ward}
  \nabla_\mu \langle T^{\mu\nu}\rangle - \langle \mathcal{O}_I \rangle
  \nabla^\nu \phi^I =0~.
\end{equation}
The constraint \eqref{review:Hamilton} is used to determine the
unknown functions $U$, $\Phi$, $M_{IJ}$, etc. These equations do not 
determine a unique solution. 
Actually, to completely fix one 
it is necessary to impose still more boundary conditions 
by comparing with the asymptotic AdS solution, as done in Sec.~\ref{perturb}.

In the original proposal, de~Boer, Verlinde and Verlinde included all
possible local terms $[S]_{2n}$ for $2n<d$ in the derivative
expansion. Fukuma, Matsuma and Sakai \cite{Fukuma00a} pointed out that
adding terms $[S]_d$ is physically irrelevant, since
it corresponds to adding total derivative terms to the anomaly, or terms
proportional to the beta function that therefore vanish at the fixed point. 
We shall show here that these terms in 
$[S]_d$ are needed to cancel spurious terms in the anomaly in the 
vicinity of the fixed point.

The terms of the Hamiltonian \eqref{review:Hamilton} stemming from the
local $[S]_{2n}$ can be split into
expressions of level $2n$, where $2n$ is the number of space-time
derivatives. Thus, one obtains for level zero
\begin{gather}
\label{review:H0}
  [\mathcal{H}]_0 = \frac{U}{2(d-1)}\left[-\frac{d}2 U +
  \frac{U}{4(d-1)} \beta^I G_{IJ} \beta^J + \frac{2(d-1)}U V \right],
  \\ 
\intertext{where $\beta^I$ denotes the holographic beta function and
  is defined as}
\label{review:beta}
  \beta^I = - G^{IJ} \frac{2(d-1)}U \partial_J U~. 
\end{gather}
Notice that this actually takes into account
only the lowest order in derivatives of the full beta function that
follows from varying the local part of $S$. This is in fact what 
appears in the Callan--Symanzik equation \cite{deBoer00a}. 
    
At level two, one finds
\begin{equation}
\label{review:H2}
  [\mathcal{H}]_2 = \frac{U}{2(d-1)} \left\{ -c_2 R 
  - \lambda_{IJ} \partial^\mu \phi^I \partial_\mu \phi^J + \nabla_\mu
  \left[ \left( 2(d-1) \partial_I \Phi +M_{IJ} \beta^J\right)
  \partial^\mu \phi^I \right] \right\},
\end{equation}
where
\begin{align}
\label{review:c2}
  c_2 &= (d-2) \Phi - \frac{2(d-1)}U +\beta^I \partial_I \Phi~,\\
\label{review:lambda}
  \lambda_{IJ} &= \frac{d-2}2 M_{IJ} +\frac{d-1}U G_{IJ} +\frac12
  \left( M_{IK} \partial_J \beta^K + M_{JK} \partial_I \beta^K +
  \partial_K M_{IJ} \beta^K \right).
\end{align}

In addition to the $[\mathcal{H}]_{2n}$ coming from the local terms in
$S$, there are terms involving the non-local $\Gamma$. The first
such term contains the anomalies,
\begin{equation}
\label{review:Hd}
  [\mathcal{H}]_\Gamma = -\frac{U}{2(d-1)} \left( \langle T\rangle +\beta^I
  \langle \mathcal{O}_I \rangle \right).
\end{equation}

The Hamiltonian constraint \eqref{review:Hamilton} is incorporated by
setting each term in $[\mathcal{H}]_{2n}$ to zero for $2n<d$ and
trying to solve this system of descent equations for the unknowns 
$U$, $\Phi$, $M_{IJ}$, etc. For $n=0$, this gives an equation for $U$
in terms of $V$, 
\begin{equation}
\label{review:Ueq}
  \frac1{2d(d-1)} \beta^I G_{IJ} \beta^J = 1 - \frac{4(d-1)V}{dU^2}~,
\end{equation}
after substituting eqn.\ \eqref{review:beta}. We must observe here
that eqn.\ \eqref{review:Ueq} does not determine $U$ uniquely. In
fact, we will revert to AdS/CFT results in order to make sure that the lowest
orders in the Taylor expansion of $U$ correspond to the leading local terms of
the on-shell action. 
As we are interested in interpolating renormalisation group trajectories, we
should also assume that $U$ has another critical point in the IR. This 
boils down to a nonperturbative boundary condition on the $U$ we are 
considering, and determines it completely \cite{Campos00}.

For $n=1$ and $d>2$, we have the three equations
\begin{align}
\label{review:level21}
  2(d-1) \partial_I \Phi + \beta^J M_{IJ} &= 0~,\\
\label{review:level22}
  c_2 &=0~,\\
\label{review:level23}
  \lambda_{IJ} &=0~. 
\end{align}
Although these are three equations for only two unknowns, the system
is not over--determined, since the equations are not functionally
independent. In fact, eqn.\ \eqref{review:level21} ensures that 
\begin{equation}
\label{review:c2lambda}
  (d-1) \partial_I c_2 = -\lambda_{IJ} \beta^J~,
\end{equation}
so that eqn.\ \eqref{review:level22} implies eqn.\ \eqref{review:level23}.

Writing down $[\mathcal{H}]_{2n}$ becomes increasingly involved for
$n>2$. We list $[\mathcal{H}]_4$ in the appendix for later use.  

If one could solve the descent equations to any level, starting from
level zero, one would obtain all counter terms $[S]_{2n}$. However,
typically, the descent equations will break at some
level $n$, in which case one must revert to $[\mathcal{H}]_{2n}
+[\mathcal{H}]_\Gamma =0$, which yields the holographic anomaly of the
boundary field theory. Moreover, the anomaly involves also the unknown
functions of $[S]_d$, which should reflect the ambiguity of choosing a
renormalisation scheme. 
We will use the freedom to add these finite terms to bring the anomaly
to a standard form close to the fixed point.

\section{Anomalies in the Vicinity of a Fixed Point}
\label{perturb}
In this section we analyse the vicinity around a given fixed point by
expansion of the descent equations of level zero to four in powers of
the scalar field. To a certain extent, this has been done in
\cite{Fukuma00a}, but we repeat the analysis here to be
self-contained.

For simplicity, we shall consider the case of a single scalar field
with $G_{II}=1$. Furthermore, without loss of generality, let us
assume that the fixed point is at 
$\phi=0$. For the holographic beta function, we shall substitute 
\begin{equation}
\label{perturb:beta}
  \beta = (\Delta-d) \phi~.
\end{equation}
We are also neglecting higher derivative terms of the full beta function, 
as discussed below eqn.\ \eqref{review:beta}.  
Generically, our notation will be 
\begin{equation}
\label{perturb:form}
  A(\phi) = A^{(0)} + A^{(1)} \phi +\frac12 A^{(2)} \phi^2 +\cdots~, 
\end{equation}
although for most quantities some of the terms are absent.

Let us start with the expansion of $U$, which is essentially the input
at the top of the descent equations. The constant $U^{(0)}$ is known
from pure gravity calculations in AdS/CFT
\cite{Liu98-1,Arutyunov98-1}, and the quadratic term follows directly
from eqns.\ \eqref{review:beta} and \eqref{perturb:beta}, while there
is no linear term, 
\begin{equation}
\label{perturb:U}
  U(\phi) = -\frac{2(d-1)}l + \frac{\Delta-d}{2l} \phi^2 +\cdots~.
\end{equation}
Thus, we have made sure that $U$ is the leading term of the on-shell
action. Now, writing the potential $V$ as 
\begin{equation}
  V(\phi) = V_0 - \frac12 m^2 \phi^2 +\cdots
\end{equation}
and setting $[\mathcal{H}]_0=0$, yields 
\begin{equation}
  V_0 = \frac{d(d-1)}l^2 \quad \text{and} \quad
  m^2 l^2 = \Delta(\Delta-d)~,\label{formula34}
\end{equation}
reproducing the AdS cosmological constant at the fixed
point and the AdS/CFT mass formula. To quadratic order in $\phi$, all
terms of the level zero constraint can be solved, so that no anomaly
is obtained at this level. This comes somewhat as a surprise, since
from AdS/CFT one should expect an anomaly proportional to $\phi^2$
for operators of dimension $\Delta=d/2$ \cite{deHaro00a}. 
However, the absence of the anomaly can be explained by the fact that
here we are dealing with bare fields.  We will show at the end of this
section how it can be recovered when translating to renormalised
quantities.

Let us proceed to the level two equations starting with the
coefficient $\lambda_{IJ}$, given in eqn.\ \eqref{review:lambda}. One finds
\begin{equation}
\label{perturb:lambda}
  \lambda = M^{(0)} \left( \Delta-\frac{d}2 -1 \right) - \frac{l}2
  +\mathcal{O}(\phi)~.
\end{equation}
Obviously, for operators of dimension $\Delta \neq d/2+1$,
$\lambda=0$ can be solved to lowest order with the result
\begin{equation}
\label{perturb:M}
  M^{(0)} = \frac{l}{2(\Delta-d/2-1)}~.
\end{equation}
In contrast, for operators of dimension $\Delta=d/2+1$, $\lambda$
contributes a term $l/2\, \partial_\mu \phi \partial^\mu \phi$ to the
anomaly, just as expected. Moreover, in this anomalous case the value
$M^{(0)}$ remains undetermined.

Given $M^{(0)}$, eqn.\ \eqref{review:level21} can always be solved and gives
\begin{equation}
\label{perturb:Phi2}
  \Phi^{(2)} = -M^{(0)} \frac{\Delta-d}{2(d-1)}~.
\end{equation}
Last, eqn.\ \eqref{review:c2} is expanded as
\begin{equation}
\label{perturb:c2}
  c_2 = (d-2) \Phi^{(0)} +l + \phi^2 \left[
  \left(\Delta-\frac{d}2-1\right) \Phi^{(2)} + \frac{(\Delta-d)l}{4(d-1)} 
  \right] +\cdots~.
\end{equation}
For operators of dimension $\Delta\neq d/2+1$, the quadratic term vanishes
by virtue of eqns.\ \eqref{perturb:Phi2} and \eqref{perturb:M}, while, for
$\Delta=d/2+1$, $c_2$ contributes a term $(d-2)l/[8(d-1)] \phi^2 R$ to
the anomaly.
Moreover, in dimensions $d\neq 2$, the constant part of $c_2$ can be
solved and gives
\begin{equation}
\label{perturb:Phi0} 
  \Phi^{(0)} = -\frac{l}{d-2}~,
\end{equation}
while, for $d=2$, we obtain the standard Weyl anomaly $-lR$. 

Our treatment of the level two descent equations 
illustrates the occurrence of anomalies in the Hamilton--Jacobi
approach to holography. In summary, we have found the following level
two anomalies in the neighbourhood of a given fixed point,
\begin{equation}
\label{perturb:anom2}
  \langle T \rangle +\beta \langle \mathcal{O}\rangle = \begin{cases}
  \frac{l}2 \partial_\mu \phi \partial^\mu \phi +
  \frac{(d-2)l}{8(d-1)} \phi^2 R +\mathcal{O}(\phi^3)~, &\text{if
  $\Delta=d/2+1$,}\\  
  -lR +\mathcal{O}(\phi^3)~, &\text{if $d=2$.}
  \end{cases}
\end{equation}

Let us now come to the anomalies in the neighbourhood of the fixed
point stemming from the level four descent equations. These will
occur, if $d=4$, or if $\Delta=d/2+2$. We have listed the level four
terms of the Hamiltonian in appendix \ref{H4}. It must be stressed
that the level four equations can only be used as they stand, if the
level two equations can be solved, \emph{i.e.}, if there is no level two
anomaly. In these cases terms like $R_{\mu\nu} \langle T^{\mu\nu}
\rangle$ contribute to level four. 
This means, that the following arguments are valid only, if
$d\neq 2$ and $\Delta\neq d/2+1$.

Starting with the term in $[\mathcal{H}]_4$ proportional to $R^2$, we
have from eqn.\ \eqref{H4:psi1} 
\begin{equation}
\label{perturb:psi1}
\begin{split}
  \psi_1 &= -(d-4) \Psi_1^{(0)} + \frac{dl}{4(d-1)} \left( \Phi^{(0)}
  \right)^2 -\phi^2 \left\{ \left( \Delta-\frac{d}2-2 \right)
  \Psi_1^{(2)} \right. \\
  &\quad \left.- \frac{dl}{4(d-1)} \left[ \frac{\Delta-d}{4(d-1)}
  \left( \Phi^{(0)} \right)^2 + \Phi^{(0)} \Phi^{(2)} \right]
  +\frac{l}2 \left( \Phi^{(2)} \right)^2 \right\} +\cdots~,
\end{split}
\end{equation}
where eqn.\ \eqref{perturb:U} has been used. Thus, if $d=4$, we cannot
solve for $\Psi_1^{(0)}$, which results in a contribution to the
anomaly, $\psi_1=l/3 \left( \Phi^{(0)} \right)^2= l^3 /12$. 
If $\Delta=d/2+2$, $\Psi_1^{(2)}$ remains undetermined, and we have
\[ \psi_1 = - \frac{(d-4)l^3\phi^2}{128(d-1)^2(d-2)^2}
(d^3-4d^2+16d-16)~. \]

Similarly, from eqn.\ \eqref{H4:psi2} we find
\begin{equation}
\label{perturb:psi2}
\begin{split}
  \psi_2 &= -(d-4) \Psi_2^{(0)} -l \left( \Phi^{(0)} \right)^2 \\
  &\quad -\phi^2 \left\{ \left( \Delta-\frac{d}2-2\right) \Psi_2^{(2)}
  +l \left[ \frac{\Delta-d}{4(d-1)} \left( \Phi^{(0)} \right)^2 +
  \Phi^{(0)} \Phi^{(2)} \right]\right\} +\cdots~.
\end{split}
\end{equation}
Again, for $d=4$, we find the contribution $\psi_2 = -l
\left(\Phi^{(0)}\right)^2 = -l^3/4$, whereas, for $\Delta=d/2+2$, we
have
\[ \psi_2 = \frac{(d-4)l^3\phi^2}{8(d-2)^2}~. \]

It is easy to see from eqn.\ \eqref{H4:psi3} that there is no
impediment to setting $\psi_3=0$ to quadratic order, which therefore
does not generate anomaly terms. Similarly, to quadratic order in
$\phi$, one can solve $\chi=\xi=\zeta=\kappa=0$. 

The next equation that generates an anomaly term is
\eqref{H4:sigma}. Indeed, we have 
\begin{equation}
\label{perturb:sigma}
  \sigma = -2 \left( \Delta-\frac{d}2-2\right) L^{(1)}\phi 
  + l M^{(0)} \Phi^{(2)} \phi +\cdots~, 
\end{equation}
so that, for $\Delta=d/2+2$, 
\[ \sigma = \frac{(d-4) l^3 \phi}{16(d-1)} + \cdots~.\]

The remaining descent equations, \eqref{H4:gamma}, \eqref{H4:nu},
\eqref{H4:alpha}, \eqref{H4:epsilon} and \eqref{H4:tau}, generate
further anomaly terms, but, in contrast to the anomaly terms obtained
so far, they are not uniquely determined. This non-uniqueness reflects
the possibility of having various renormalisation schemes, and the
results should be physically equivalent. In the following, we shall
give one possible choice. Let us start by setting eqn.\ \eqref{H4:tau}
to zero, which is achieved at lowest order in $\phi$ by $(d-2) N^{(0)}
= 4(d-1) K^{(0)}$. Then, from eqn.\ \eqref{H4:epsilon}, we have 
\begin{equation}
\label{perturb:epsilon}
  \epsilon = -2 \left( \Delta- \frac{d}2 -2\right) A^{(0)} -\frac{l}2
  M^{(0)}+\cdots~, 
\end{equation}
so that, for $\Delta=d/2+2$, $\epsilon=-l^3/8$. 

Next, eqn.\ \eqref{H4:gamma} becomes
\begin{equation}
\label{perturb:gamma}
  \gamma = -2 \left( \Delta- \frac{d}2 -2\right) K^{(0)} -
  \frac{(d-2)l}{4(d-1)} \Phi^{(0)} M^{(0)} +\cdots~, 
\end{equation}
and the contribution to the anomaly for $\Delta=d/2+2$ is
$\gamma=l^3/[8(d-1)]$. 

Finally, there is no difficulty to set either $\nu$ or $\alpha$ to
zero to lowest order in $\phi$, and we shall choose $\alpha=0$. Then,
eqns.\ \eqref{H4:nu} and \eqref{H4:alpha} yield
\begin{equation}
\label{perturb:nu}
  \nu = \phi \left[ 2\left(\Delta-\frac{d}2-2\right) N^{(0)} +l
  \Phi^{(0)} M^{(0)} \right] +\cdots~.
\end{equation}
Hence, for $\Delta=d/2+2$, we find $\nu = -\phi l^3/[2(d-2)]$.

Let us now summarise the level four anomalies. For $d=4$, we have 
\begin{equation}
\label{perturb:anom4d}
  \langle T \rangle +\beta \langle \mathcal{O}\rangle = \frac{l^3}4
  \left( \frac13 R^2 -R_{\mu\nu}R^{\mu\nu} \right)
  +\mathcal{O}(\phi^3)~,
\end{equation}
whereas, collecting all the terms for $\Delta=d/2+2$, we have
\begin{equation}
\label{perturb:anom4}
\begin{split}
  \langle T \rangle +\beta \langle \mathcal{O}\rangle &= - \frac{l^3
  \phi^2 (d-4)}{8(d-2)^2} \left( \frac{d^3-4d^2+16d-16}{16 (d-1)^2}
  R^2 - R_{\mu\nu} R^{\mu\nu} \right) \\
  &\quad + \frac{l^3(d-4)}{16(d-1)} R \phi\nabla^2 \phi +
  \frac{l^3}{8(d-1)} R \partial_\mu \phi \partial^\mu \phi \\
  &\quad - \frac{l^3}8 \nabla^2 \phi \nabla^2 \phi
  - \frac{l^3}{2(d-2)} \left( R^{\mu\nu} -\frac12 g^{\mu\nu} R \right)
  \phi \nabla_\mu \partial_\nu \phi +\mathcal{O}(\phi^3)~.
\end{split}
\end{equation}

Next, let us translate the bare anomalies
\eqref{perturb:anom2}, \eqref{perturb:anom4d} and
\eqref{perturb:anom4} to renormalised ones, where the asymptotic
fields $\hat{g}_{\mu\nu}$ and $\hat{\phi}$ are the field theory
sources.
This amounts to taking into account the Jacobian $\mathcal{A}(x,r)$ discussed 
around eqn.\ \eqref{jacob}. As the anomaly is homogeneous in derivatives, the
derivative corrections will contribute in higher orders and can therefore 
be neglected
here. Notice that had we had level two 
anomalies, they 
would have contributed  on level four through the first correction in 
$\mathcal{A}$. 
In a general anomaly higher order derivative terms do appear 
together with appropriately suppressed powers of $\e{-r/l}$, thus not 
contributing to the fixed point result.    

What will be important, however, is the $r$-dependent scaling in 
${\cal A}$. Writing 
\begin{align}
\label{perurb:Ohat}
  \langle\hat{\mathcal{O}} \rangle &= \frac1{\sqrt{\hat{g}}} \frac{\delta
  \Gamma}{\delta \hat{\phi}}~, &
\langle \hat{T}^{\mu\nu} \rangle &= \frac2{\sqrt{\hat{g}}} \frac{\delta
  \Gamma}{\delta \hat{g}_{\mu\nu}}~, 
\end{align}
we find, using eqns.\ \eqref{review:asym1} and \eqref{review:asym2}, 
\begin{align}
\label{perturb:TThat}
  \langle T \rangle &= \e{-dr/l} \langle \hat{T} \rangle +\cdots~,\\
\label{perturb:OOhat}
  \langle \mathcal{O} \rangle &= \e{-\Delta r/l} 
  \langle\hat{\mathcal{O}} \rangle +\cdots~.
\end{align}
Moreover, 
\begin{equation}
\label{perturb:betahat}
  \beta = (\Delta-d) \e{(\Delta-d)r/l} \hat{\phi} +\cdots =
  \e{(\Delta-d)r/l} \hat{\beta} +\cdots~,
\end{equation}
so that 
\begin{equation}
\label{perturb:TOTOhat}
  \langle T \rangle + \beta \langle \mathcal{O} \rangle = 
  \e{-dr/l} \left( \langle \hat{T} \rangle + \hat{\beta}
  \langle\hat{\mathcal{O}} \rangle \right) + \cdots~.
\end{equation}
It is easy to see that the right hand sides of eqns.\
\eqref{perturb:anom2}, \eqref{perturb:anom4d} and
\eqref{perturb:anom4} scale in the same manner when written in terms
of the hatted fields. Therefore, our results directly translate into
the anomalies of the boundary field theories with properly identified
sources. 

Finally, let us return to the $\Delta=d/2$ anomaly mentioned earlier below 
eqn.\ (\ref{formula34}). In
this case, the leading asymptotic behaviour of the field $\phi$ is not
given by eqn.\ \eqref{review:asym2}, but rather by 
\begin{equation}
\label{perturb:Deltad2}
  \phi(x,r) = r \e{-dr/(2l)} \hat{\phi} (x) +\cdots
\end{equation}
as follows from the equations of motion in the asymptotic AdS background.
Hence, 
\[ \dot{\phi} = \e{-dr/(2l)} \hat{\phi} + \frac{(\Delta-d)}{l} \phi +\cdots~. \]
This behaviour should be described by the flow equation
\eqref{review:HJ1}, which means that the momentum $\pi_I$ must receive 
non-negligible contributions from
the one-point function, $\pi_I = \partial_I U +\langle \mathcal{O}_I
\rangle +\cdots$, which gives 
\[ \langle \mathcal{O} \rangle = \e{-dr/(2l)} \hat{\phi} +\cdots~.\]
Thus, after going to renormalised quantities, we find the level zero
contribution 
\begin{equation}
\label{perturb:Deltad2anom}
  \langle \hat{T}\rangle+\hat{\beta} \langle \hat{\mathcal{O}} 
  \rangle = -\frac{d}2
  \hat{\phi}^2 +\mathcal{O}(\hat{\phi}^3)~,
\end{equation}
which is the anomaly of the expected form.

\section{Discussion}
\label{global}
We have found that dBVV's Hamilton--Jacobi approach to
the holographic renormalisation group can be successfully applied in
order to derive anomalies in the vicinity of a
given fixed point. Regarding the boundary data at the cut-off as the
bare sources 
of the boundary field theory, the anomalies are formally obtained,
except in the case $\Delta=d/2$, and the results are easily
translated to the proper renormalised sources. 

We should 
point out that our method of renormalising the sources differs from
that of dBVV's in that they considered the boundary data at some
renormalisation scale $r_R$ as the renormalised sources \cite{deBoer00a}. 
In contrast, we, in agreement with the standard method for calculating 
holographic correlation functions 
\cite{Witten98-1,Gubser98-1,Freedman99a,Arutyunov00a}, 
consider the asymptotically rescaled boundary
data---hatted fields in eqns.\ \eqref{review:asym1} and
\eqref{review:asym2}---as the renormalised sources. This renders the
interpretation of the descent equations rather difficult, if the cut-off
boundary data are far away from the UV fixed point. In other words, even if
we can write down an expansion for the bulk--boundary propagator ${\cal A}$ near
a UV fixed point, the full expression would be needed for a nonperturbative
analysis of the interpolating trajectory.

In this paper we have observed that the assumption that the flow of 
the bulk fields is essentially determined by the leading local term
$[S]_0$---implicit in eqn.\ \eqref{review:beta}---becomes invalid. 
We have seen at the end of the last section that
a non-local term had to be taken into account for operators
$\Delta=d/2$ even close to the fixed point. Moreover, there is no guarantee
that one can solve the descent equations up to an arbitrary level
$2n$. Consider, for example, operators of dimension $\Delta=3$ in $d=4$. 
Here, the level
two equations are broken already close to the UV fixed point. This
means that there is a level two anomaly, which contributes via terms
like $R_{\mu\nu} \langle T^{\mu\nu} \rangle$ to the level four
equations.  

The level two descent equations seem to be the only ones which can be
analysed globally. If one takes the point of view that the cut-off
boundary data are the bare sources of the boundary field theory at the
cut-off, one is able to identify a class of $c$-functions in $d=2$. 
Consider again the level two terms of the Hamiltonian, eqn.\
\eqref{review:H2}, and $d=2$. 
There is no obvious difficulty to set the total
derivative term to zero globally, \emph{i.e.}, we let
\begin{gather}
\label{global:Phi}
  -2 \partial_I \Phi = \beta^J M_{IJ}~,
\intertext{and obtain the anomaly}
\label{global:anom2}
 \langle T\rangle +\beta^I \langle \mathcal{O}_I \rangle  
 = -c_2 R - \lambda_{IJ} \partial_\mu \phi^I \partial^\mu
 \phi^J+\cdots~,
\end{gather}
where the ellipses stand for terms with more derivatives or
curvatures, which would arise from higher levels.
As shown in Sec.\ \ref{perturb}, setting $M_{IJ}=\Phi=0$ gives
unexpected anomalies in the vicinity of a fixed point except for
marginal operators ($\Delta=d=2$). 
The coefficients $c_2$ and $\lambda_{IJ}$ are functions of $\phi$ at
the cut-off. Moreover, eqn.\ \eqref{global:Phi} ensures that 
eqn.\ \eqref{review:c2lambda} is still valid, \emph{i.e.}, 
\begin{equation}
\label{global:c2lambda}
  \partial_I c_2 = -\lambda_{IJ}\beta^J~.
\end{equation}
It is interesting to notice that the flow-derivative of $c_2$ is given by 
\begin{equation}
\dot c_2 = \frac{U}2 \beta^I\beta^J \lambda_{IJ}~. 
\end{equation}
Therefore assuming $\lambda_{IJ}$ positive definite, $c_2$ is monotonous and 
can actually be interpreted as a $c$-function of the holographic
renormalisation group.

The presence of an anomaly means that the functions 
$M_{IJ}$ and $\Phi$ cannot
be adjusted as to make $c_2$ and $\lambda_{IJ}$ vanish identically. 
We may therefore try to use the freedom, already seen in 
Sec.~\ref{perturb}, to choose these
functions in such a way that $c_2$ becomes monotonous. Nevertheless we
must also make sure that $\lambda_{IJ}=0$ at the fixed points without
marginal operators. 
Generically, then, there will be ambiguities in the definition of 
the $c$-function, as also expected from the field theory analysis 
\cite{Anselmi00a}.

On the other hand, starting from the UV fixed point, for non-marginal 
operators ($\Delta\neq 2$ ) one could also attempt to solve $\lambda_{IJ}=0$
globally. Then it would follow from eqn.\ \eqref{global:c2lambda} that 
$c_2$ is constant. This would be a ``non-renormalisation'' theorem for
the $c$-function as it would be independent of both the cut-off and the 
couplings. 
However, since this would lead to an $M_{IJ}$ that is singular at the
IR fixed point, one should conclude that the perturbative
renormalisation scheme we are using cannot be extrapolated towards the
IR, and therefore a global choice of a regular $M_{IJ}$ is not possible.

It would be interesting to compare our results, where the sources are 
the asymptotically $r$-independent parts of the coupling, to those 
of  dBVV, where the relationship between bare and renormalised 
quantities is somewhat different \cite{deBoer00a}---there 
for a given field configuration the field at the 
cut-off is the source. Here asymptotically rescaled
fields are cut-off independent for any given field configuration, but 
the cut-off boundary is still needed to regularise the on-shell action. 
This radial behaviour appears to be responsible for some of the expected 
QFT features observed above. 
But, we leave the investigation of the meaning of this open for further
studies.

\section*{Acknowledgements}
We would like to thank M.~Bertolini, M.~Bianchi, A.~Karch, A.~Miemiec, 
A.~C.~Petkou, and H.~S.~Reall for discussions and useful comments. 
We also thank the 
organisers of the RTN conference ``Latest developments in M-Theory'', 
Paris, February 1--8, 2001, during which this work was essentially 
completed. 
The work was supported by the European Commission RTN programme
HPRN-CT-2000-00131, in which J.~K.\ is associated with Imperial
College London and W.~M.\ with INFN, Sezione di Frascati. D.~M.\
acknowledges partial support from PPARC through SPG\#613.

\begin{appendix}

\section{Level Four Hamiltonian}
\label{H4}
Here, we give a summary of the level four terms in the Hamiltonian. 
The local terms $[S]_4$ are 
\begin{equation}
\label{H4:S4}
\begin{split}
  [S]_4 &= \int d^d x\sqrt{g} \left[ \Psi_1 R^2 
  + \Psi_2 R_{\mu\nu}R^{\mu\nu} 
  + \Psi_3 R_{\mu\nu\lambda\rho} R^{\mu\nu\lambda\rho}
  + L_I R \nabla^2 \phi^I 
  \phantom{\frac12} \right. \\
  &\quad + K_{IJ} R \partial_\mu \phi^I \partial^\mu \phi^J 
  + N_{IJ} \left( R^{\mu\nu} -\frac12 g^{\mu\nu} R
  \right) \partial_\mu \phi^I \partial_\nu \phi^J 
  + A_{IJ} \nabla^2 \phi^I \nabla^2 \phi^J \\
  &\left. \phantom{\frac12} 
  + B_{IJK} \nabla_\mu \nabla_\nu \phi^I \partial^\mu \phi^J
  \partial^\nu \phi^K 
  + C_{IJKL} \partial_\mu \phi^I \partial^\mu \phi^J \partial_\nu
  \phi^K \partial^\nu \phi^L \right]. 
\end{split}
\end{equation}
According to the general procedure, this leads to 
\begin{align}
\label{H4:H4}
  [\mathcal{H}]_4 &= \frac{U}{2(d-1)} \left( -2[\pi]_4 -
  \beta^I [\pi_I]_4 \right) + [\pi_{\mu\nu}]_2 [\pi^{\mu\nu}]_2
  -\frac1{d-1} \left([\pi]_2\right)^2 +\frac12 [\pi_I]_2 G^{IJ} [\pi_J]_2 \\
\label{H4:H4expl}
\begin{split}
  &= \frac{U}{2(d-1)} \left[ R^2 \psi_1
  + R_{\mu\nu}R^{\mu\nu} \psi_2 
  + R_{\mu\nu\lambda\rho} R^{\mu\nu\lambda\rho} \psi_3 
  + \nabla^2 (R \chi)
  + \nabla_\mu (R\partial^\mu \phi^I \xi_I) \phantom{\frac12} \right.\\
  &\quad +\nabla^2 (\nabla^2 \phi^I \zeta_I) 
  + \nabla_\mu (\nabla^2 \phi^I \partial^\mu \phi^J \kappa_{IJ})
  + R\nabla^2 \phi^I \sigma_I + R \partial_\mu \phi^I \partial^\mu
  \phi^J \gamma_{IJ} \\
  &\quad + \left(R^{\mu\nu}-\frac12 g^{\mu\nu} R \right) \nabla_\mu
  \partial_\nu \phi^I \nu_I 
  + \left(R^{\mu\nu}-\frac12 g^{\mu\nu} R \right) \partial_\mu \phi^I
  \partial_\nu \phi^J \alpha_{IJ} \\
  &\quad + \nabla^2 \phi^I \nabla^2 \phi^J \epsilon_{IJ}
  + \nabla_\mu\partial_\nu \phi^I \nabla^\mu \partial^\nu \phi^J
  \tau_{IJ} 
  + \nabla^2 \phi^I \partial_\mu \phi^J
  \partial^\mu \phi^K \rho_{IJK}\\
  &\left. \phantom{\frac12}
  + \nabla_\mu \partial_\nu \phi^I \partial^\mu \phi^J \partial^\nu
  \phi^K \theta_{IJK}
  + \partial_\mu \phi^I \partial^\mu \phi^J \partial_\nu \phi^K
  \partial^\nu \phi^L \omega_{IJKL} \right].
\end{split}
\end{align}
The coefficients in eqn.\ \eqref{H4:H4expl} are given by 
\begin{align}
\label{H4:psi1}
  \psi_1 &= -(d-4) \Psi_1 -\beta^I \partial_I \Psi_1 -\frac{d}{2U}
  \Phi^2 +\frac{d-1}U \partial_I \Phi G^{IJ} \partial_J \Phi, \\
\label{H4:psi2}
  \psi_2 &= -(d-4) \Psi_2 -\beta^I \partial_I \Psi_2 +\frac{2(d-1)}{U}
  \Phi^2, \\
\label{H4:psi3}
  \psi_3 &= -(d-4) \Psi_3 -\beta^I \partial_I \Psi_3, \\
\label{H4:chi}
  \chi &= 4(d-1) \Psi_1 + d \Psi_2 + 4 \Psi_3 - \beta^I L_I, \\
\label{H4:xi}
  \xi_I &= (d-2) L_I +2 L_J \partial_I \beta^J + 2 K_{IJ} \beta^J, \\
\label{H4:zeta}
  \zeta_I &= 2(d-1) L_I - 2 A_{IJ} \beta^J, \\
\label{H4:kappa}
  \kappa_{IJ} &= 4(d-1) K_{IJ} +2 (d-2) A_{IJ} +4 A_{IK} \partial_J
  \beta^K -2 \beta^K (B_{KIJ} - B_{IJK}),\\
\label{H4:sigma}
  \sigma_I &= -(d-4) L_I -\beta^J \partial_J L_I - L_J \partial_I
  \beta^J -\frac{2(d-1)}U M_{IJ} G^{JK} \partial_K \Phi, \\
\label{H4:gamma}
\begin{split}
  \gamma_{IJ} &= - \frac{d-2}2
  N_{IJ} + (d+2) K_{IJ} -\beta^K \partial_K K_{IJ} - K_{IK} \partial_J
  \beta^K - K_{JK} \partial_I \beta^K \\
  &\quad - L_K \partial_I \partial_J \beta^K +\frac12 \beta^K (B_{IJK}
  +B_{JIK} -B_{KIJ}) +\frac{d-2}{2U} \Phi M_{IJ} \\
  &\quad - \frac{d-1}{U} \partial_L \Phi G^{LK} (\partial_J
  M_{IK} +\partial_I M_{JK} -\partial_K M_{IJ}),  
\end{split} \\
\label{H4:nu}
  \nu_I &= 2(d-2) \partial_I \Psi_2 +8 \partial_I \Psi_3 + 2 \beta^J
  N_{IJ} -\frac{4(d-1)}U \Phi \partial_I \Phi,\\
\label{H4:alpha}
\begin{split}
  \alpha_{IJ} &= 2(d-2) \partial_I \partial_J \Psi_2 + 8 \partial_I
  \partial_J \Psi_3 -2 (d-3) N_{IJ} +4 (d-1) K_{IJ} \\
  &\quad + \beta^K ( \partial_I N_{JK} +\partial_J N_{IK} - \partial_K
  N_{IJ} + B_{IJK} +B_{JIK} -B_{KIJ} ) \\
  &\quad + \frac{2(d-1)}U \Phi ( M_{IJ}
  -2 \partial_I \partial_J \Phi) ,
\end{split} \\
\begin{split}
\label{H4:epsilon}
  \epsilon_{IJ} &= (d-2) N_{IJ} -4(d-1) K_{IJ} -(d-4) A_{IJ} - \beta^K
  \partial_K A_{IJ} - A_{IK} \partial_J \beta^K - A_{JK} \partial_I
  \beta^K \\
  &\quad - \beta^K ( B_{IJK} +B_{JIK} -B_{KIJ} ) -\frac{2(d-1)}U
  \partial_I \Phi \partial_J \Phi + \frac{d-1}U M_{IK} G^{KL} M_{LJ} ,
\end{split}\\
\label{H4:tau}
  \tau_{IJ} &= -(d-2) N_{IJ} + 4(d-1) K_{IJ} + 
  \beta^K ( B_{IJK} +B_{JIK} -B_{KIJ} ) + \frac{2(d-1)}U
  \partial_I \Phi \partial_J \Phi .
\end{align}
The remaining three coefficients $\rho_{IJK}$, $\theta_{IJK}$ and
$\omega_{IJKL}$ will not be needed. 

\end{appendix}

\end{document}